\documentclass[notitlepage,a4paper,aps,prd,onecolumn,superscriptaddress,nofootinbib,groupedaddress]{revtex4}

\usepackage{enumerate}
\usepackage{amsmath}
\usepackage{amsfonts}
\usepackage{amssymb}
\usepackage[utf8]{inputenc}
\usepackage[T1]{fontenc}
\usepackage{mathtools}
\usepackage{wasysym}
\usepackage{accents}
\newcommand{\lc}[1]{\accentset{\circ}{#1}}%Levi-Civita connection
\newcounter{subeq}

\DeclareMathAlphabet{\mathds}{U}{BOONDOX-ds}{m}{n}
\usepackage[dvipsnames]{xcolor}

\usepackage[colorlinks=true]{hyperref}
\usepackage{amsthm}

\theoremstyle{definition}

\theoremstyle{plain}

\newcommand{\dd}{\mathrm{d}}

\allowdisplaybreaks

\begin{document}
\title{General Teleparallel Modifications of Schwarzschild Geometry}

\author{Sebastian Bahamonde}
\email{sbahamonde@ut.ee}
\affiliation{Laboratory of Theoretical Physics, Institute of Physics, University of Tartu, W. Ostwaldi 1, 50411 Tartu, Estonia.}
\affiliation{Laboratory for Theoretical Cosmology, Tomsk State University of
Control Systems and Radioelectronics, 634050 Tomsk, Russia (TUSUR)}

\author{Christian Pfeifer}
\email{christian.pfeifer@ut.ee}
\affiliation{Laboratory of Theoretical Physics, Institute of Physics, University of Tartu, W. Ostwaldi 1, 50411 Tartu, Estonia.}

\begin{abstract}
Teleparallel theories of gravity are described in terms of the tetrad of a metric and a flat connection with torsion. In this paper, we study spherical symmetry in a modified teleparallel theory of gravity which is based on an arbitrary function of the five possible scalars constructed from the irreducible parts of torsion. This theory is a generalisation of the so-called New General Relativity theory. We find that only two scalars are different to zero in spherical symmetry and we solve the corresponding field equations analytically for conformal Teleparallel gravity, and then perturbatively around Schwarzschild geometry for the general perturbative theory around GR. Finally we compute phenomenological effects from the perturbed solutions such as the photon sphere, perihelion shift, Shapiro delay and the light deflection. We find their correspondent correction to the standard GR contribution and their dependence on the three model parameters.
\end{abstract}

\maketitle

%%%%%%%%%%%%%%%%%%%%%%%%%%%%%%%%%%%%%%%%%%%%%%%%%%%%%%%
\section{Introduction}\label{sec:intro}
General Relativity (GR) is a very successful theory which describes phenomena from the motion of planetary system, such as the solar system, via gravitational waves from binary systems, to the evolution of the accelerating expanding Universe as whole~\cite{misner1973gravitation,will_2018} to high precision. Over the last years many access to many new observables have been achieved either astrophysical sources or cosmology. The most recent examples are the observation of the shadow of the black hole at the center of the galaxy M87~\cite{Akiyama:2019cqa} and the detection of gravitational waves \cite{Abbott:2016blz,TheLIGOScientific:2017qsa,Abbott:2016nmj,Abbott:2017vtc}. All of these observations are so far mostly consistent with GR.

Nevertheless, there are some theoretical and observational problems that GR faces and for which an explanation within GR is missing. Most prominent are the dark energy and dark matter phenomenology as well as the question about the nature of the cosmological constant, the emergence of singularities and the issue of a missing theory of quantum gravity~\cite{Bertone:2004pz,Riess:1998cb,Perlmutter:1998np,Peebles:2002gy,Copeland:2006wr,Weinberg:1988cp,Kiefer:2005uk,Bodendorfer:2016uat}. Additionally, the growing tensions between cosmological parameters have emerged after recent new data sets had been evaluated, such as the so-called $H_0$ or $\sigma_8$ tensions~\cite{Aghanim:2018eyx,Riess:2018uxu,Wong:2019kwg}. Also, in the realm of gravitational wave observations exist new intriguing result, such as the one found in GW190814, suggesting that a neutron star can have a mass within the so-called mass gap that is predicted by GR~\cite{Abbott:2020khf}. These observations together demonstrate that pieces in our understanding of gravity are still missing and that GR might not be the final answer.

Approaches to extend and modify GR, to obtain an understanding of the discussed problems are numerously discussed and proposed in the literature proposed~\cite{Clifton:2011jh,Heisenberg:2018vsk,Ishak:2018his,Nojiri:2017ncd,Capozziello:2011et}. The most straightforward generalisation is to keep Riemannian geometry as geometry of spacetime, as in GR, and either extend the Lagrangian by generalising the Einstein-Hilbert action, or to add new degrees of freedom through additional gravitational scalar, vector or tensor fields, and couple these to further curvature scalars then the Ricci scalar. Many such theories have been constructed containing advantages and disadvantages compared to GR.

Another route for modified gravity is to modify geometry of spacetime which represents the gravitational field. Instead of Riemannian geometry one may consider an affine geometry base on a  metric and a dynamical independent affine connection \cite{Hehl:1999sb,Hehl:1994ue}. Depending on the properties the connection has, non-metricity, torsion and or curvature, different models can be constructed and are investigated \cite{BeltranJimenez:2019tjy,Jarv:2018bgs,Conroy:2017yln,Olmo:2011uz}. In the context of quantum gravity phenomenology and the standard model extension one leaves the realm of metric affine gravity and considers curved velocity and momentum spaces \cite{AmelinoCamelia:2008qg,Tasson:2016xib,Liberati:2013xla}, for example in terms of Finsler or Hamiltonian geometry \cite{Pfeifer:2019wus,Hohmann:2019sni,Barcaroli:2015xda,Relancio:2020zok}.

In this article we consider teleparallel gravity, which is based on so-called Teleparallel geometry, where the connection on the spacetime manifold has torsion, is metric compatible and has no curvature, As fundamental variable one uses the tetrad of the metric insted of the metric itself~\cite{Aldrovandi:2013wha,Weitzenbock1923}. On the basis of teleparallel geometry on can simply reformulate GR as ``Teleparallel equivalent of General Relativity" (TEGR)~\cite{Aldrovandi:2013wha} and starting from thereon, one can construct modified teleparallel theories of gravity~\cite{Cai:2015emx,Maluf:2013gaa,Bahamonde:2016kba,Bahamonde:2015zma,Hohmann:2017duq,Bahamonde:2019shr}. 

The TEGR Lagrangian is constructed from the so-called torsion scalar $\mathbb{T}$, which is a specific combination of contractions of the torsion tensor. It turns out that the Ricci scalar computed with the Levi-Civita connection is related to the scalar torsion as $\mathring{R}=-\mathbb{T}+B$, and then, the TEGR action differs from the Einstein-Hilbert action by a boundary term $B$. Therefore, the TEGR field equations are identical to the Einstein's field equations.

The first TEGR modification was formulated in~\cite{Hayashi:1979qx}, where the authors generalised the TEGR Lagrangian by constructing the three non-parity violating scalars that one can construct form the torsion tensor and combined these linearly with arbitrary coefficients weighing them. A very specific numerical choice of these coefficients yields TEGR itself. This theory was labelled as ``New General Relativity". In~\cite{Jimenez:2019tkx} it was found that the unique non-pathological theory around Minkowski is the family of parameters which recovers TEGR. 

Later, several authors~\cite{Ferraro:2006jd,Bengochea:2008gz,Krssak:2015oua} studied a generalised theory considering a Lagrangian with a arbitrary function $f(\mathbb{T})$ of the torsion scalar $T$. This theory has became very popular in the last decade, with different types of studies such as cosmology~\cite{Bengochea:2008gz,Cai:2011tc,Bamba:2010wb,Dent:2011zz} and astrophysics~\cite{Ferraro:2011ks,Nunes:2019bjq,Bamba:2013ooa,Cai:2018rzd,Nunes:2018evm,Farrugia:2018gyz,Ahmed:2016cuy,Boehmer:2011gw,Hohmann:2018jso}. One important issue about this theory is the possibility of being strongly coupled which has been suggested by recent papers~\cite{Jimenez:2020ofm,Golovnev:2018wbh}. Basically, these authors have shown that the new degrees of freedom of $f(\mathbb{T})$~\cite{Blagojevic:2020dyq,Ferraro:2020tqk} do not show up in both Minkowski and FLRW backgrounds, and then one cannot trust perturbation techniques around these space-time geometries. Due to this, some other modified Teleparallel theories have been proposed  in the literature.  For example, in~\cite{Bahamonde:2015zma} so-called $f(T,B)$ gravity was formulated, where the boundary term $B$ connecting the torsion scalar with the Ricci scalar is considered in the Lagrangian. So far, no relevant interesting exact spherically symmetric vacuum solutions exist in this model ~\cite{Bahamonde:2019jkf}. However, what exists are perturbation solutions which investigate first order teleparallel perturbations of TEGR and find perturbative solutions around the famous Schwarzschild solution of GR~\cite{Bahamonde:2019zea,Bahamonde:2020bbc,Ruggiero:2015oka,DeBenedictis:2016aze}.

Based on the ideas of $f(\mathbb{T})$ gravity and New General Relativity, a class of theories has been suggested which is based on an arbitrary function of all the possible five, parity even and parity odd, scalars that one can construct from torsion~\cite{Bahamonde:2017wwk}. The investigation of the phenomenology and viability of these theories is an ongoing project in the literature. It is already known that the FLRW background of these theories are identical to the $f(\mathbb{T})$, but its perturbations have not been studied yet. 

In this paper, we will study this theory in spherical symmetry and find the most general perturbative solution around TEGRs Schwarzschild solution. We will study how these solutions affect the motion of particles and derive the  observables: circular photon orbits, deflection of light, Shapiro delay and the perihelion shift. With this we extend the existing studies of these observables based on weak $f(\mathbb{T})$ gravity. 

This paper is organised as follows: In Sec.~\ref{sec:TPSph} we give a short introduction to Teleparallel gravity and also we present the most general action constructed from the contraction of torsion tensor (up to quadratic contractions). In this section, we also discuss how one determines the most general spherically symmetric tetrad in Weitzenb\"ock gauge, find the field equation in spherical symmetry and obtain exact and perturb solutions. Sec.~\ref{sec:pheno} is devoted to studying particle motion phenomenology of these spherically symmetric solutions, where we compute photon sphere, perihelion shift, Shapiro delay and the light deflection. We conclude our main results in~\ref{sec:conclusion}.

Throughout the paper we denote Latin (Greek) indices to refer to tangent(space-time) space. The tetrad and its inverse are denoted by $h^{a}{}_{\mu}$ and $h_{a}{}^{\mu}$, quantities with overcircle on top $\circ$ denote that they are computed from the Levi-Civita connection, and our signature convention is  $(+,-,-,-)$. We also work in the units where $G=c=1$.

%%%%%%%%%%%%%%%%%%%%%%%%%%%%%%%%%%%%%%%%%%%%%%%%%%%%%%%
\section{Teleparallel theories of gravity in spherical symmetry}\label{sec:TPSph}
To analyse spherically symmetric teleparallel theories of gravity we study them in terms of the most general spherically symmetric Weitzenb\"ock tetrad. We use the decomposition of the torsion tensor into tensorial, vectorial and axial part to construct the five canonical quadratic torsion scalars, and find that only two of these are non-vanishing. With this finding we are able to display the spherically symmetric field equations for most classes of teleparallel gravity theories considered in the literature.

%%%%%%%%%%%%%%%%%%%%%%%%%%%%%%%%%%%%%%%%%%%%%%%%%%%%%%%
\subsection{Teleparallel gravity}\label{ssec:TPGrav}
Teleparallel theories of gravity have a long history in physics \cite{Aldrovandi:2013wha,Ferraro:2006jd,Krssak:2018ywd}. They are formulated in terms of the tetrads $h^a = h^a{}_\mu \dd x^\mu$ of a metric $g = \eta_{ab}h^a{}_\mu h^b{}_\nu \dd x^\mu \otimes \dd x^\nu$, their duals $h_a = h_a{}^\mu \partial_\mu$ and a flat, metric compatible spin connection $\omega^a{}_b = \omega^a{}_{b\mu}\dd x^\mu$ with torsion. The flatness and metric compatibility demand on the spin connection yield, that its components are generated by local Lorentz transformation matrices $\Lambda^a{}_b$ as $\omega^a{}_{b\mu} = \Lambda^a{}_c \partial_\mu (\Lambda^{-1})^c{}_{b}$. The torsion of the spin connection is given by
\begin{align}
    	T^a{}_{\mu\nu} = 2 \left(\partial_{[\mu}h^a{}_{\nu]} + \omega^a{}_{b[\mu} h^b{}_{\nu]}\right)\,.
\end{align}

Due to the specific form of the spin connection, it is always possible to introduce the so called Weitzenb\"ock tetrad $h^a_W = h^b (\Lambda^{-1})^a{}_b$ for which the torsion becomes
\begin{align}
    T^a_W{}_{\mu\nu} =  2 \partial_{[\mu}h^a_W{}_{\nu]}\,.
\end{align}
It has been shown in detail in the literature that it is equivalent to study teleparallel theories of gravity either, with a tetrad and a spin connection and looking for solutions of the field equations, or, to simply consider the Weitzenb\"ock tetrad and a vanishing spin connection. In the latter case the tetrad has to solve the symmetric and the anti-symmetric part of the field equations \cite{Golovnev:2017dox,Hohmann:2017duq,Krssak:2015oua}.

In what follows we will always work with the Weitzenb\"ock tetrad and vanishing spin connection. For the sake of readability we drop the label ${}_W$ on the geometric objects. Moreover we will work with the pure spacetime index torsion 
\begin{align}
    T^{\sigma}{}_{\mu\nu} = h_a{}^{\sigma}T^a{}_{\mu\nu}\,.
\end{align}
that is related to the mixed index torsion by a contraction with an inverse tetrad.
To construct an action for teleparallel theories of gravity, one uses the torsion tensor as fundamental building block. The simplest scalars one can construct require contractions between two torsion tensors. There are five independent such scalars, which can be constructed in a most systematic way by decomposing the torsion tensor into the so called vector, axial and tensor torsion, see~\cite{Bahamonde:2017wwk}, 
\begin{align}
v_{\mu} = T^{\nu}{}_{\nu\mu}, \quad a_{\mu} = \frac{1}{6}\epsilon_{\mu\nu\rho\sigma}T^{\nu\rho\sigma},\quad 
t_{\mu\nu\rho} = T_{(\mu\nu)\rho} + \frac{1}{3}\left(T^{\sigma}{}_{\sigma(\mu}g_{\nu)\rho} - T^{\sigma}{}_{\sigma\rho}g_{\mu\nu}\right)\,.
\end{align}
The three parity even torsion scalars are
\begin{align}\label{eq:tseven}
	T_{\rm vec} = v_{\mu}{v}^{\mu},\quad  T_{\rm ax} = a_{\mu}{a}^{\mu},\quad T_{\rm ten} = t_{\lambda\mu\nu}{t}^{\lambda\mu\nu}\,,
\end{align} 
while the two parity odd ones are
\begin{align}\label{eq:tsodd}
	P_{1} = v_{\mu}a^{\mu}, \quad 
	P_{2} = \epsilon_{\mu\nu\rho\sigma}t_{\lambda}{}^{\mu\nu}t^{\lambda\rho\sigma} \,.
\end{align}
The most general action which can be constructed from these terms is~\cite{Bahamonde:2017wwk}
\begin{align}\label{eq:fTTTTTact}
    \tilde S[h,\Psi] = \int_M d^4x\,h \left[\frac{1}{2\kappa^2} f(T_{\rm ten},T_{\rm vec},T_{\rm ax}, P_1, P_2) + \mathcal{L}_{\rm m}(g,\Psi)\right]\,,
\end{align}
where $h=\textrm{det}(h^a{}_\mu)=\sqrt{-g}$ and $\mathcal{L}_{\rm m}(g,\Psi)$ is the matter field Lagrangian. The matter fields are minimally coupled to the tetrads via the metric they generate. Surely, to obtain a well defined action integral it is important that the the function $f$ is chosen such that the parity odd terms appear in a way that they combine into terms that are parity even in total, i.e.\ they must appear multiplied with each other in an even power.

To avoid the complication when parity odd terms are included, the literature focused on $f(T_{\rm ten},T_{\rm vec},T_{\rm ax})$-theories, which were introduced in \cite{Bahamonde:2017wwk},
\begin{equation}\label{eq:fTTTact}
    S[h,\Psi]=\int d^4x \, h\, \left[\frac{1}{2\kappa^2}f(T_{\rm ten},T_{\rm vec},T_{\rm ax})+\mathcal{L}_{\rm m}(g,\Psi)\right]\,.
\end{equation}
A most famous class of teleparallel gravity theories, which fit in the framework just introduced, are the \emph{new general relativity} theories \cite{Hayashi}. They are defined by the most general Lagrangian that is linear in the parity even torsion scalars, parametrized by three constants $c_t$, $c_v$ and $c_a$
\begin{align}
    f = c_t T_{\rm ten} + c_v T_{\rm vec} + c_a T_{\rm ax}\,.
\end{align}
Fixing the values of the constants to $c_t=\frac{2}{3}$, $c_v=-\frac{2}{3}$, $c_a = \frac{3}{2}$ the Lagrangian becomes the torsion scalar
\begin{align}\label{eq:tegr}
    \mathbb{T} = \frac{2}{3} T_{\rm ten} - \frac{2}{3} T_{\rm vec} + \frac{3}{2} T_{\rm ax}
\end{align}
that is, up to a total divergence, i.e\ a boundary term in the metric, identical to the Ricci scalar of the metric induced by the tetrads. It defines the teleparallel equivalent of general relativity (TEGR) \cite{Aldrovandi:2013wha,Maluf:2013gaa} which has the same equations as General Relativity.

In the following we recall the most general spherically symmetric tetrad in Weitzenb\"ock gauge and study static perturbations of TEGR in spherical symmetry, to identify the phenomenological imprints of teleparallel theories of gravity.

%%%%%%%%%%%%%%%%%%%%%%%%%%%%%%%%%%%%%%%%%%%%%%%%%%%%%%%
\subsection{Spherical symmetry}\label{sec:sphTPG}
To study symmetric solutions in of the field equations of a teleparallel theory of gravity, the tetrad and the spin connection, both, have to satisfy the symmetry conditions. In mathematical precise words, both of them have to be invariant under a certain set of diffeomorphism of spacetime.

The details, how one can construct such symmetric teleparallel geometries have been detailed in \cite{Hohmann:2019nat}. Here we recall the results for spherical symmetry.

Symmetries are generated by vector fields $Z_\zeta = Z_\zeta^\mu(x) \partial_\mu$ on spacetime, and, the tetrad and the spin connection are invariant under the flow of these vector fields if they satisfy the equations:
\begin{align}\label{eq:tpsymm}
\mathcal{L}_{Z_\zeta}h^{a}\,_{\mu}=-\,\lambda^{a}_{\zeta}{}_{b}h^{b}\,_{\mu} \,,\quad 
\mathcal{L}_{Z_\zeta}\omega^{a}\,_{b\mu}=\partial_{\mu}\lambda^{a}_{\zeta}{}_{b}+\omega^{a}\,_{c\mu}\lambda^{c}_{\zeta}{}_{b}-\omega^{c}\,_{b\mu}\lambda^{a}_{\zeta}{}_{c}\,.
\end{align}
These equations are the teleparallel generalisations of the well known killing equation for the metric in pseudo-Riemannian geometry. The quantity $\lambda^{a}_{\zeta}{}_{b}$ appearing here defines a Lie algebra homomorphism which maps the symmetry algebra generated by the symmetry vector fields $Z_\zeta$ into the Lorentz algebra. These are needed to impose the symmetry of the tetrad and the spin connection consistently in every Lorentz frame.

Solving these equations in the Weitzenb\"ock gauge, i.e.\ with vanishing spin connection, implies that the Lie algebra homomorphism $\lambda$ cannot depend on spacetime points, $\partial_{\mu}\lambda^{a}_{\zeta}{}_{b}=0$. Hence, the only remaining equation which needs to be solved is the symmetry condition for the tetrad for a fixed choice of $\lambda$.

To study spherically symmetric systems we consider coordinates $(t,r,\phi,\theta)$ and impose the symmetry generators
\begin{align}
    Z_1 &= \sin \phi \partial_\theta + \frac{\cos \phi}{\tan \theta}\partial_\phi\\
    Z_2 &= -\cos \phi \partial_\theta + \frac{\sin \phi}{\tan \theta}\partial_\phi\\
    Z_3 &= \partial_\phi \,,
\end{align}
which represent the $so(3)$ algebra. The only way how to map this algebra into the Lorentz algebra with a constant Lie algebra homomorphism in 4 dimensions is via the identifications
\begin{align}
    \lambda(Z_1) = 
    \left(
    \begin{array}{cccc}
    0 & 0 & 0 & 0\\
    0 & 0 & 0 & 0\\
    0 & 0 & 0 & -1\\
    0 & 0 & 1 & 0\\
    \end{array}
    \right),\quad
    \lambda(Z_2) = 
    \left(
    \begin{array}{cccc}
    0 & 0 & 0 & 0\\
    0 & 0 & 0 & 1\\
    0 & 0 & 0 & 0\\
    0 & -1 & 0 & 0\\
    \end{array}
    \right),\quad
    \lambda(Z_3) = 
    \left(
    \begin{array}{cccc}
    0 & 0 & 0 & 0\\
    0 & 0 & -1 & 0\\
    0 & 1 & 0 & 0\\
    0 & 0 & 0 & 0\\
    \end{array}
    \right)\,.
\end{align}

Solving the remaining teleparallel symmetry condition \eqref{eq:tpsymm} for the tetrad yields 
\begin{equation}
h^a{}_{\nu}=\left(
\begin{array}{cccc}
C_1 & C_2 & 0 & 0 \\
C_3 \sin\theta \cos\phi & C_4 \sin\theta \cos\phi & C_5 \cos\theta \cos \phi - C_6 \sin\phi  & -\sin\theta (C_5 \sin\phi + C_6 \cos\theta \cos\phi)  \\
C_3 \sin\theta \sin\phi & C_4 \sin\theta \sin\phi & C_5 \cos\theta \sin \phi + C_6 \cos\phi  & \sin\theta (C_5 \cos\phi - C_6 \cos\theta \sin\phi) \\
C_3 \cos\theta & C_3 \cos\theta & - C_5 \sin\theta & C_6 \sin^2\theta\\
\end{array}
\right)\label{sphtetrad}\,,
\end{equation}
where the six free functions $C_I$ ($I=1,..6$), in general, depend on $(t,r)$. The spacetime metric determined by this tetrad is
\begin{align}
    ds^2 = (C_1^2-C_3^2 ) \,dt^2 +2 (C_1 C_2-C_3 C_4 )\, dt\, dr  - (C_4^2 - C_2^2) \,dr^2 - (C_5^2 + C_6^2)(d\theta^2 + \,\sin^2 \theta d\phi^2)\,.
\end{align}
We can use the freedom of coordinate transformations to diagonalize the $dt\,dr$ part of the metric 
\begin{align}\label{eq:coordcond1}
  C_3 C_4 - C_1 C_2 = 0\,,  
\end{align}
and to set
\begin{align}\label{eq:coordcond2}
    (C_5^2 + C_6^2) = C^2 \Leftrightarrow C_5^2 = C^2 - C_6^2\,. 
\end{align}
Thus, one of the six free functions of the tetrad can be fixed by coordinate choice and a second can be directly related to a component of the metric.

To summarize, the most general spherically symmetric Weitzenb\"ock tetrad, i.e.\ with vanishing spin connection, which defines a spherically symmetric teleparallel geometry contains five free functions and is given by \eqref{sphtetrad}, in which one of the six appearing free functions is fixed by the coordinate choice conditions \eqref{eq:coordcond1} and one can directly be expressed in terms of one metric component \eqref{eq:coordcond2}.

The fundamental torsion scalars $T_{\rm ten},T_{\rm vec}$ and $T_{\rm ax}$, which are the building blocks of the theories we are interested in, are rather lengthy expressions for the most general spherically symmetric tetrad. That is why we display them in the appendix~\ref{appendixB}.

An important remark is that for Minkowski spacetime, none of the scalars are vanishing.

So far our analysis was based only on symmetry considerations. It is clear that these spherically symmetric tetrads include the ones which were found in the context of solving the antsymmetric field equations in $f(\mathbb{T})$-gravity \cite{Bohmer:2011si}.

Next we search for solutions of the $f(T_{\rm ten},T_{\rm vec},T_{\rm ax})$-gravity field equations in spherical symmetry.

%%%%%%%%%%%%%%%%%%%%%%%%%%%%%%%%%%%%%%%%%%%%%%%%%%%%%%%
\subsection{Field equations}\label{sec:FEQsphTPG}
We seek to determine the five free functions in the tetrad \eqref{sphtetrad} such that they solve the field equations for $f(T_{\rm ten},T_{\rm vec},T_{\rm ax})$ gravity. These are obtained by variation of the action \eqref{eq:fTTTact} with respect to the tetrad components
\begin{align}
E_\beta{}^\mu= &-2 f_{T_{\rm vec}} \left(T^\mu{}_{\beta \sigma}  v^\sigma+ v^\mu  v_\beta \right) - 2h^a{}_\beta\, \lc{\nabla}_\nu \left[  f_{T_{\rm vec}}  (v^\mu h_a{}^{\nu }   -v^\nu  h_a{}^{\mu }  )\right] \nonumber\\
&-\frac{2}{3}  f_{T_{\rm ax}} \epsilon_{\sigma \alpha}{}^{\mu\lambda} a^\sigma  T^{\alpha}{}_{\beta\lambda}  -\frac{2}{3} h^a{}_\beta\lc{\nabla}_\nu \left(f_{T_{\rm ax}} \epsilon_{\sigma\alpha}{}^{\nu\mu}h_{a}{}^\alpha    a^\sigma\right) \nonumber\\
&-f_{T_{\rm ten}}\Big(2T^\rho{}_{\beta\sigma}T_{\rho}{}^{\mu \sigma} + T^\mu{}_{\rho \sigma}T^{\rho}{}_{\beta}{}^{\sigma} +
T^\alpha{}_{\rho \beta} T^{\rho}{}_{\alpha}{}^{\mu} - T^\mu{}_{\beta \sigma} v^\sigma    
- v^\mu  v_\beta \Big)\nonumber\\
&+ h^a{}_\beta\lc{\nabla}_\nu \Bigl[ f_{T_{\rm ten}} \left(
2 T_a{}^{\mu\nu} - T^{\mu\nu}{}_{a}+T^{\nu\mu}{}_{a} + v^\mu h_a{}^{\nu} - v^\nu h_a{}^{\mu}\right)\Bigr] +f\delta^\mu_\beta=2\kappa^2h^{a}{}_\beta\Theta_{a}{}^{\mu}\,,
\end{align}
where the energy-momentum tensor was defined as
\begin{equation}
\Theta_a{}^{\mu} = \frac{1}{h} \frac{\delta (h\mathcal{L}_{\rm m})}{\delta h^a{}_{\mu}} \,. 
\label{Theta}
\end{equation}
The original derivation of these equations can be found in \cite{Bahamonde:2017wwk}.

We start by considering the antisymmetric part of the field equations which determine at least two of the five free functions, before we determine the remaining three free functions from the symmetric field equations.

\subsubsection{The antisymmetric equations}
There are two non-vanishing antisymmetric field equations $E_{[tr]},E_{[\theta\phi]}$ for the tetrad \eqref{sphtetrad}. Using~\eqref{eq:coordcond1} and \eqref{eq:coordcond2}, they can be written in the following implicit form
\begin{eqnarray}
   E_{[tr]}&=&Q_{1}f_{\rm vec}+Q_{2}f_{\rm ax}+\Big(Q_{1}-\frac{9}{4}Q_2\Big)f_{\rm ten}+Q_3f'_{\rm vec}+\Big(\frac{3 C_1 \left(C_3 C'_1-C_1 C'_3\right)}{2 C_4 \left(C_1^2-C_3^2\right)^2}-\frac{1}{2}Q_3\Big)f'_{\rm ten}=0\,,\label{anti1}\\
  E_{[\theta\phi]}&=&\sin\theta  \left(\csc^2\theta +1\right)\Big[Q_4f_{\rm vec}+Q_5 f_{\rm ax}+\Big(Q_4-\frac{9}{4}Q_5\Big)f_{\rm ten}+Q_6f'_{\rm ax}-\frac{3}{2}\Big(\frac{C_1^2 \sqrt{C^2-C_5^2}}{C^2 C_4 \left(C_1^2-C_3^2\right)}+\frac{3}{2}Q_6\Big)\Big]=0\,, \label{anti2}
\end{eqnarray}
where primes denote differentiation with respect to $r$ and $Q_i$ are functions of $r$, that are explicitly written in the appendix~\ref{appendixC}. The antisymmetric field equations can be solved in different ways but since we do not want to constrain our model, i.e., the form of $f$, we must set
\begin{eqnarray}
Q_1=Q_2=Q_4=Q_5=0\,,  \label{sol1}
\end{eqnarray}
leading to the following solution
\begin{eqnarray}
    C_5(r)=\pm\,C(r)\,,\quad C_3(r)=0\,.
\end{eqnarray}
There is another branch that solves~\eqref{sol1} but this one constrains more than three functions and then, the corresponding metric will not have all the degrees of freedom.

Using these findings from the antisymmetric field equations in the coordinate condition \eqref{eq:coordcond1} we can still choose $C_1=0$ or $C_2=0$. From the $(+,-,-,-)$ sign convention of the metric we identify the choice $C_2=0$ as suitable.

\subsubsection{The symmetric field equations}
From the antisymmetric field equations we concluded that the solution, without restricting the theory or the metric, is given by the Weitzenb\"ock tetrads of the form 
\begin{equation}
h^a{}_{\nu}=\left(
\begin{array}{cccc}
\sqrt{A} & 0 & 0 & 0 \\
0 & \sqrt{B} \cos\phi \sin\theta & C \cos\phi \cos\theta  & -C \sin\phi \sin\theta  \\
0 & \sqrt{B} \sin\phi \sin\theta  & C \sin\phi \cos\theta  & C \cos\phi \sin\theta \\
0 & \sqrt{B} \cos\theta & -C \sin\theta & 0 \\
\end{array}
\right)\label{tetrad}\,,
\end{equation}
where for convenience we renamed $C_1 = \sqrt{A}$ and $C_4 = \sqrt{B}$. This tetrad yields the static spherically symmetric metric
\begin{equation}
ds^2=A(r) \,dt^2-B(r)\, dr^2-C(r)^2(d\theta^2+\sin^2\theta d\phi^2)\,,\label{metric}
\end{equation}
and the five fundamental torsion scalars in~\eqref{Tscalar1}-\eqref{Tscalar5} simplify to
\begin{align}
    T_{\rm ten}  &= - \frac{(C A' + 2 A (\sqrt{B} - C'))^2}{4 A^2 B C^2}\,,\\
    T_{\rm vec} &= -\frac{1}{B}\left( \frac{A'}{2 A} + \frac{2}{C} (C' - \sqrt{B}) \right)^2\,,\\
    T_{\rm ax} &= P_1 = P_2 = 0\,. 
\end{align}
This in particular also implies that in $f(T_{\rm ten}, T_{\rm vec}, T_{\rm ax}, P_1, P_2)$ theories, the parity odd terms $P_1$ and $P_2$ have no influence in spherical symmetry. This is true since they always appear at least as square or mutual product between them in the Lagrangian, by the fact that we consider parity even Lagrangians, and thus at least they appear linearly in the field equations.

The symmetric field equations of $f(T_{\rm ten}, T_{\rm vec}, T_{\rm ax})$-gravity for the tetrad \eqref{tetrad} are
\begin{align}
\frac{1}{2} \kappa ^2 \rho&=\frac{f_{T_{\rm ten}} }{8 r^2 A^2 B^2}\left[-r^2 B A'^2+r A \left(-r A' B'+2 B^{3/2} A'+2 B \left(r A''+A'\right)\right)+2 A^2 (r B'+2 B^{3/2}-2 B)\right]\nonumber\\
&+\frac{f_{T_{\rm vec}} }{8 r^2 A^2 B^2}\left[-r^2 B A'^2+r A \left(-r A' B'-4 B^{3/2} A'+2 B (r A''+4 A')\right)-4 A^2 (r B'+2 B^{3/2}-2 B)\right]\nonumber\\
&+\frac{1}{4 r A B}\Big[r A' \left(f'_{T_{\rm ten}}+f'_{T_{\rm vec}}\right)+2 A (\sqrt{B}-1) \left(f'_{T_{\rm ten}}-2 f'_{T_{\rm vec}}\right)\Big]+\frac{1}{4} f \,,\label{Eq1}\\
\frac{1}{2} \kappa ^2 p_{\rm r}&= -\frac{f_{T_{\rm ten}} }{8 r^2 A^2 B}\left(r A'-2 A\right) \Big[r A'+2 A (\sqrt{B}-1)\Big]-\frac{f_{T_{\rm vec}}}{8 r^2 A^2 B} \left(r A'+4 A\right) \left[r A'-4 A (\sqrt{B}-1)\right] - \frac{1}{4} f\,,\label{Eq2}\\
\frac{1}{2} \kappa ^2 p_{\rm l}&=-\frac{f_{T_{\rm ten}} }{16 r^2 A^2 B^2}\left[r^2 B A'^2+r A \left(r A' B'-2 B \left(r A''+A'\right)\right)+A^2(-2 r B'-8 B^{3/2}+4 B^2+4 B)\right]\nonumber\\
&+\frac{f_{T_{\rm vec}} }{8 r^2 A^2 B^2}\left[r^2 B A'^2+r A \left(r A' B'+6 B^{3/2} A'-2 B \left(r A''+4 A'\right)\right)+A^2 (4 r B'+16 B^{3/2}-8 B^2-8 B)\right]\nonumber\\
&+\frac{f'_{T_{\rm ten}} }{8 r A B}\left[r A'+2 A (\sqrt{B}-1)\right]+\frac{f'_{T_{\rm vec}} }{4 B}\left[\frac{4 (\sqrt{B}-1)}{r}-\frac{A'}{A}\right]-\frac{1}{4} f\,.\label{Eq3}
\end{align}
Here, $f_{T_{\rm ten}}=\partial f/\partial T_{\rm ten}$, $f_{T_{\rm vec}}=\partial f/\partial T_{\rm vec}$ and we have assumed an anisotropic fluid for the matter whose energy density is $\rho$ and its lateral and radial pressures $p_{\rm l}$ and $p_{\rm r}$, respectively. For $f = \mathbb{T}$, see \eqref{eq:tegr}, we have that $f_{T_{\rm ten}} = f_{T_{\rm vec}} = f_{T_{\rm ax}} = const.$ and thus $f'_{T_{\rm ten}} = f'_{T_{\rm vec}} = f'_{T_{\rm ax}} = 0$. For this choice the above equations are identical to the Einstein equations for a spherically symmetric metric \eqref{metric} with $C=r$. Their unique vacuum solution then is the Schwarzschild solution $A = B^{-1} = 1 - 2 M/r$. Consequently, choosing $f = f(\mathbb{T}) = f((2/3) T_{\rm ten}-(2/3)  T_{\rm vec})$, which gives $f_{T_{\rm ten}}=-f_{T_{\rm vec}}=(2/3)f_{\mathbb{T}}(\mathbb{T})$, on recovers correctly the spherically symmetric $f(\mathbb{T})$ field equations, which were for example reported in~\cite{Bahamonde:2019zea}.

Next, we will solve these equations for different choices of the function $f(T_{\rm ten}, T_{\rm vec}) = f(T_{\rm ten}, T_{\rm vec},0,0,0)$. In particular we find the influence of weak teleparallel perturbations on Schwarzschild geometry. 

%%%%%%%%%%%%%%%%%%%%%%%%%%%%%%%%%%%%%%%%%%%%%%%%%%%%%%%
\subsection{Solving the symmetric field equations} 
In general it is very difficult to find non-perturbative solutions for the field equations \eqref{Eq1} to \eqref{Eq3}. Exceptions are the TEGR case when $f = \mathbb{T}$, and, as we will see below, conformal teleparallel gravity.

To investigate the effects of teleparallel modifications of general relativity  in spherically symmetry we will will solve the field equations for theories that are perturbations of TEGR.

%%%%%%%%%%%%%%%%%%%%%%%%%%%%%%%%%%%%%%%%%%%%%%%%%%%%%%
\subsubsection{Conformal teleparallel gravity} 
Conformal teleparallel gravity is introduced in the literature in terms of specifying $f$ to, see~\cite{Maluf:2011kf},
\begin{equation}
    f_{\rm conf}(T_{\rm ax},T_{\rm ten},T_{\rm vec})=\frac{9}{4}\alpha \hat{T}^2=\frac{9}{4}\alpha \Big(\frac{3}{2} T_{\rm ax}+\frac{2}{3} T_{\rm ten}\Big)^2\,,
\end{equation} 

In spherical symmetry the function reduces to
\begin{align}
    f_{\rm conf}(T_{\rm ten},T_{\rm vec}) = \alpha T^2_{\rm ten} = \alpha \frac{(r A' + 2 A (\sqrt{B} - 1))^4}{16 A^4 B^2 r^4} \,.
\end{align}

Plugging this function into the field equations we find that
\begin{equation}\label{eq:Bconf}
B=\frac{9 \left(r A'-2 A\right)^2}{4 A^2}\,, 
\end{equation}
solves them identically, i.e.\ $A$ is not determined. Choosing the Ansatz $A = \frac{1}{B}$ yields 
\begin{equation}
    A(r)=\frac{1}{9}\Big(K_0 r-1\Big)^2\,.
\end{equation}

Our little example here proves that this conformal teleparallel theory of gravity is not predictive, since it does not determine the components of a spherically symmetric metric without further assumptions.

%%%%%%%%%%%%%%%%%%%%%%%%%%%%%%%%%%%%%%%%%%%%%%%%%%%%%%%%%%%%%%%%%%%%%%%%%%%%%%%%%%%%%%%%%%%%%%%%%%%%%
\subsubsection{Perturbations around Schwarzschild geometry} \label{sec:weak}
To study the influence of teleparallel modifications of general relativity to particle motion in spherical symmetry we consider $f$ as a power law
\begin{align}
    f(T_{\rm ten},T_{\rm vec},T_{\rm ax}, P_1, P_2) 
    &=  \sum_{I,J,K,L,M} \frac{1}{(I+J+K+L+N)!}D_{IJKLN} T_{\rm ten}^I T_{\rm vec}^J T_{\rm ax}^K P_1^L P_2^N\\
    &=\Lambda + c_t T_{\rm ten} + c_v T_{\rm vec} + c_a T_{\rm ax} + \tfrac{1}{2}(\alpha_1 T_{\rm ten}^2 + \alpha_2 T_{\rm vec}^2 + 2 \alpha_3 T_{\rm ten} T_{\rm vec})\nonumber \\
    &+\tfrac{1}{2} (\alpha_4 T_{\rm ax}^2 + \alpha_5 T_{\rm ax} T_{\rm ten} + \alpha_6 T_{\rm ax} T_{\rm vec} + \alpha_7 P_1^2 + \alpha_8 P_2^2 + \alpha_9 P_1 P_2 + ...)\,,
\end{align}
where the sum of the integers $L+N$ must always be even to obtain a parity even Lagrangian. We called the first order coefficients $D_{10000} = c_t$, $D_{01000} = c_v$ and $C_{00100} = c_a$ and the second order coefficients $D_{20000} = \alpha_1$,  $D_{02000} = \alpha_2$,  $D_{11000} = 2 \alpha_3$, $D_{00200} = \alpha_4$, $D_{10100} = \alpha_5$, $D_{01100} = \alpha_6$, $D_{00020} = \alpha_7$, $D_{00002} = \alpha_8$ and $D_{00011} = \alpha_9$. The zeroth order is a cosmological constant term and the first order defines the new general relativity Lagrangian. 

Any teleparallel theory of gravity which is built from a function of the type $f(T_{\rm ten},T_{\rm vec},T_{\rm ax}, P_1, P_2)$, and these are nearly all discussed models in the literature, admits such an expansion for a small values of the zeroth order, i.e.\ for example as perturbation around Minkowski spacetime as a solution, or, in asymptotic flat or weak field regime regions. This expansion is a straightforward generalisation of the the expansion of $f(\mathbb{T})$ models. The later are contained in the above expansion by setting $c_t=2/3$, $c_v=-2/3$, $c_a =3/2$.

According to the discussion in the previous section the above expansion reduces in spherical symmetry to
\begin{align}
    f(T_{\rm ten},T_{\rm vec}) = f(T_{\rm ten},T_{\rm vec},0, 0, 0) 
    &=  \sum_{I,J,K,L,M} \frac{1}{(I+J+K+L+N)!}D_{IJKLN} T_{\rm ten}^I T_{\rm vec}^J T_{\rm ax}^K P_1^L P_2^N\\
    &=\Lambda + c_t T_{\rm ten} + c_v T_{\rm vec} + \frac{1}{2}(\alpha_1 T_{\rm ten}^2 + \alpha_2 T_{\rm vec}^2 + 2 \alpha_3 T_{\rm ten} T_{\rm vec}) + ...\,.
\end{align}

To derive the influence of teleparallel correction to Schwarzschild geometry we choose the coefficients $c_t=\frac{2}{3}$, $c_v=-\frac{2}{3}$ and introduce a perturbation bookkeeping parameter $\epsilon$ to define the family of weak teleparallel perturbations of TEGR
\begin{align}\label{eq:weakfT}
    f_{weak}(T_{\rm ten},T_{\rm vec}) =\Lambda + \frac{2}{3} (T_{\rm ten} - T_{\rm vec}) + \frac{\epsilon}{2}(\alpha_1 T_{\rm ten}^2 + \alpha_2 T_{\rm vec}^2 + 2 \alpha_3 T_{\rm ten} T_{\rm vec}) + \mathcal{O}(\epsilon^2)\,,
\end{align}
for which we derive solutions of the field equations to first order in $\epsilon$ of the form
\begin{align}
A(r)&=1-\frac{2M}{r}+\epsilon \, a(r)+ \mathcal{O}(\epsilon^2)\,,\\  
B(r)&= \left(1-\frac{2M}{r} \right)^{-1}+\epsilon \,  b(r)+ \mathcal{O}(\epsilon^2)\,.\label{expansion}
\end{align}
In earlier works weak $f(\mathbb{T})$ gravity has been considered, i.e. a Lagrangian of the type $f = \mathbb{T} + \frac{1}{2}\epsilon \alpha \mathbb{T}^2$. This theory is contained in our more general approach here by setting $\alpha_1=\alpha_2=-\alpha_3=(4/9)\alpha$.

In what follows all expression are understood that they are derived up to first order in $\epsilon$ and we drop the higher order symbol $\mathcal{O}(\epsilon^2)$.

Using \eqref{eq:weakfT} and \eqref{expansion} in the field equations~\eqref{Eq1}-\eqref{Eq3}, and performing an expansion up to first order in $\epsilon$, one obtains
\begin{eqnarray}
2\kappa ^2\epsilon \rho&=&\epsilon  \Big[\mu^4 \left(\frac{2 \left(r b'-b\right)}{r^2}-\frac{81 (7 \alpha_1-9 \alpha_2-2 \alpha_3)}{32 r^4}\right)+\mu^2 \left(\frac{4 b}{r^2}+\frac{9 (11 \alpha_1+91 \alpha_2+2 \alpha_3)}{8 r^4}\right)+\frac{9 \mu^3 (3 \alpha_1-9 \alpha_2-\alpha_3)}{r^4}\nonumber\\
&&-\frac{3 \mu (9 \alpha_1+16 \alpha_2+\alpha_3)}{r^4}+\frac{3 (\alpha_1+\alpha_2-\alpha_3)}{r^4 \mu}+\frac{3 (\alpha_1+9 \alpha_2-2 \alpha_3)}{8 r^4 \mu^2}+\frac{\alpha_1-2 \alpha_2-\alpha_3}{r^4 \mu^3}+\frac{9 (\alpha_1+\alpha_2+2 \alpha_3)}{32 r^4 \mu^4}\nonumber\\
&&+\frac{11 \alpha_1-13 \alpha_2+142 \alpha_3}{16 r^4}\Big] \,,\label{Eq1b}\\
-2\kappa^2 \epsilon p_{\rm r}
&=& \epsilon \Big[-\frac{32 r^3 a'+32 r^2 a+\alpha_1+241 \alpha_2-190 \alpha_3}{16 r^4}+\frac{16 r^2 a+3 \alpha_1-33 \alpha_2+6 \alpha_3}{8 r^4 \mu^2}-\frac{243 \mu^4 (\alpha_1+\alpha_2+2 \alpha_3)}{32 r^4}\nonumber\\
&&+\mu^2 \left(\frac{2 b}{r^2}+\frac{9 (3 \alpha_1-33 \alpha_2-26 \alpha_3)}{8 r^4}\right)+\frac{27 \mu^3 (\alpha_1+2 \alpha_2+3 \alpha_3)}{2 r^4}-\frac{9 \mu (3 \alpha_1-6 \alpha_2+\alpha_3)}{2 r^4}\nonumber\\
&&+\frac{9 (\alpha_1+2 \alpha_2-\alpha_3)}{2 r^4 \mu}+\frac{-\alpha_1+2 \alpha_2+\alpha_3}{2 r^4 \mu^3}-\frac{3 (\alpha_1+\alpha_2+2 \alpha_3)}{32 r^4 \mu^4}\Big]\,,\label{Eq2b}\\
-2\kappa^2 \epsilon p_{\rm l}
&=& \epsilon \Big[\frac{4 r^3 a'-3 \alpha_1+3 \alpha_2}{8 r^4 \mu^2}-\frac{16 r^4a''+24 r^3 a'-8 r^2 a+5 \alpha_1+557 \alpha_2+562 \alpha_3-8 r^2 b}{16 r^4}\nonumber\\
&&+\frac{\mu^2 \left(9 (-7 \alpha_1+67 \alpha_2+28 \alpha_3)+4 r^3 b'\right)}{8 r^4}+\frac{\mu^4 \left(81 (5 \alpha_1+9 \alpha_2+14 \alpha_3)+16 r^3 b'-16 r^2 b\right)}{32 r^4}\nonumber\\
&&+\frac{-16 r^2 a-3 \alpha_1+9 \alpha_2+6 \alpha_3}{32 r^4 \mu^4}-\frac{9 \mu^3 (9 \alpha_1+33 \alpha_2+37 \alpha_3)}{4 r^4}+\frac{3 \mu (27 \alpha_1-7 \alpha_2+65 \alpha_3)}{4 r^4}\nonumber\\
&&+\frac{-15 \alpha_1+69 \alpha_2+9 \alpha_3}{4 r^4 \mu}-\frac{\alpha_1+7 \alpha_2-\alpha_3}{4 r^4 \mu^3}\Big]\,,\label{Eq3b}
\end{eqnarray}
where $\mu=(1-2M/r)^{1/2}$. One can solve these equations straightforwardly for vacuum case $\rho=p_{\rm r}=p_{\rm l}=0$ and find the metric component functions 
\begin{eqnarray}
a(r)&=&\frac{1}{r^2 \left(\mu^2-1\right)^2}\Big[\frac{1}{320} (387 \alpha_1+659 \alpha_2-514 \alpha_3)+\frac{81}{64} \mu^8 (\alpha_1+\alpha_2+2 \alpha_3)-\frac{27}{10} \mu^7 (\alpha_1+2 \alpha_2+3 \alpha_3)\nonumber\\
&&-\frac{9}{64} \mu^6 (9 \alpha_1-39 \alpha_2-22 \alpha_3)+\frac{27}{5} \mu^5 (\alpha_1+2 \alpha_2+3 \alpha_3)+\frac{1}{8} \mu^4 (-11 \alpha_1-293 \alpha_2-169 \alpha_3)\nonumber\\
&&+\frac{1}{80} \mu^2 (-239 \alpha_1-2123 \alpha_2+608 \alpha_3)+\mu (\alpha_1-2 \alpha_2-\alpha_3)+\frac{-\alpha_1+2 \alpha_2+\alpha_3}{2 \mu}-\frac{3 (\alpha_1+\alpha_2+2 \alpha_3)}{64 \mu^2}\nonumber\\
&&+\log (\mu) \left(\frac{3}{8} (\alpha_1+9 \alpha_2-2 \alpha_3)-\frac{9}{8} \mu^2 (\alpha_1+9 \alpha_2-2 \alpha_3)\right)+2 (25 \alpha_2+\alpha_3) \mu^3\Big]\,,\label{sola}\\
b(r)&=&\frac{1}{r^2 \left(\mu^2-1\right)^2}\Big[\frac{9}{32} (11 \alpha_1+91 \alpha_2+2 \alpha_3)-\frac{1}{64} 27 \mu^2 (7 \alpha_1-9 \alpha_2-2 \alpha_3)+\frac{9}{5} \mu (3 \alpha_1-9 \alpha_2-\alpha_3)\nonumber\\
&&+\frac{11 \alpha_1-13 \alpha_2+142 \alpha_3}{32 \mu^2}+\frac{3 (\alpha_1+\alpha_2-\alpha_3)}{\mu^3}+\frac{201 \alpha_1-263 \alpha_2-122 \alpha_3}{160 \mu^4}+\frac{-\alpha_1+2 \alpha_2+\alpha_3}{\mu^5}-\frac{9 (\alpha_1+\alpha_2+2 \alpha_3)}{64 \mu^6}\nonumber\\
&&+\frac{-9 \alpha_1-16 \alpha_2-\alpha_3}{\mu}+\frac{3 (\alpha_1+9 \alpha_2-2 \alpha_3) \log (\mu)}{8 \mu^4}\Big]\,.\label{solb}
\end{eqnarray}
Notice that we have set the integration constants which appear in the system in such a way that we recover the standard Schwarzschild at the weak field limit at $r\rightarrow \infty$, i.e., $A\sim 1-2M/r+\mathcal{O}(1/r^2)$ and $ B\sim 1+2M/r+\mathcal{O}(1/r^2)$. One can check that if $\alpha_1=\alpha_2=-\alpha_3=(4/9)\alpha$, one recovers the squared power-law $f(\mathbb{T})=\mathbb{T}+(1/2)\mathbb{T}^2$ solution found previously in~\cite{Bahamonde:2019zea,DeBenedictis:2016aze}. The above perturbed solution is asymptotically flat since $a,b\rightarrow 0$ for $r\rightarrow \infty$ ($\mu\rightarrow 1$). 

It is worth mentioning that, similarly as it happens in $f(\mathbb{T})$ gravity, there are no first order perturbative solutions around Minkowski background ($M=0$). In this case, only if one expands the equations up to fourth order in $\epsilon$, one finds the first non-trivial corrections.

Having found the teleparallel perturbations of Schwarzschild geometry we next derive explicitly the effects on observables from point particle motion.

%%%%%%%%%%%%%%%%%%%%%%%%%%%%%%%%%%%%%%%%%%%%%%%%%%%%%%%%%%%%%%%%%%%%%%%%%%%%%%%%%%%%%%%%%%%%%%%%%%%%%
\section{Particle motion phenomenology}\label{sec:pheno}
Some classical observables related to the propagation of particles in a spherically symmetric gravitational field are the photon sphere of a black hole, which determines its shadow, the perihelion shift, the Shapiro time delay and the deflection angle of light.

The shadow of a black hole has recently been observed for the first time \cite{Akiyama:2019cqa}. Even though realistic black holes will be rotating and a derivation of the photon regions in axial symmetry is necessary, the calculation of the photon sphere in spherical symmetry is the step towards this goal.

The other three observables can be measured to high precision in the solar system. In total thus we derive four observables which can be used to constrain the parameters $\alpha_1, \alpha_2$ and $\alpha_3$ of the teleparallel perturbations.

%%%%%%%%%%%%%%%%%%%%%%%%%%%%%%%%%%%%%%%%%%%%%%%%%%%%%%%%%%%%%%%%%%%%%%%%%%%%%%%%%%%%%%%%%%%%%%%%%%%%%
\subsection{Geodesic equation and effective potential}
The following is a quick summary how to obtain the relevant equations for the desired observables. These are mainly textbook calculations for which details can be found for example in the books~\cite{Weinberg:1972kfs,Carroll:2004st,Misner:1974qy}. 

The geodesic equation, which governs the motion of point particles can be derived as Euler Lagrange equation $\frac{d}{d\tau} \dot \partial_\mu \mathcal{L} - \partial_\mu L = 0$ of the Lagrangian 
\begin{equation}
2\mathcal{L}=g_{\mu\nu}\dot{q}^{\mu}\dot{q}^{\nu}=A\,\dot{t}^2-B\,\dot{r}^2-r^2\dot{\theta}^2-r^2\sin^2{\theta}\dot{\phi}^2\,.\label{EL}
\end{equation}
The $\theta$ equation is immediately solved by setting $\theta = \pi/2$, and the $t$- and the-$\phi$ equation each reveal the existence of the constants of motion, the energy $k$, and the angular momentum $h$ 
\begin{align}\label{kh}
    k &= \frac{\partial \mathcal{L}}{\partial \dot t} =A\dot{t}= \left(1-\frac{2 M}{r}+ \epsilon\, a(r)\right)\dot{t}\,,\\
	h &= -\frac{\partial \mathcal{L}}{\partial \dot \phi} = r^2 \dot \phi\,.
\end{align}
Since physical particles must obey the normalisation condition $2 \mathcal{L} = \sigma$, where $\sigma =1$ for massive and $\sigma=0$ for massless particles, the normalisation condition yields the equation which governs the radial motion of the particles
\begin{align}\label{17}
\dot{r}^2=B^{-1}\Big(\frac{k^2}{A}-\frac{ h^2}{r^2}-\sigma\Big)\,.
\end{align}
In terms of the effective potential $V(r)= - \frac{1}{2} B^{-1}\Big(\frac{k^2}{A}-\frac{ h^2}{r^2}-\sigma\Big)$ it takes the form
\begin{align}
	\dot r^2 + 2V(r)=0\,.\label{potentialEq}
\end{align}
To first order in $\epsilon$, using \eqref{expansion}, the potential can be expanded as
\begin{align}\label{eq:pot1}
	V(r) 
	&=- \frac{1}{2} k^2 + \frac{1}{2} \left(1-\frac{2M}{r}\right) \left( \frac{h^2}{r^2} + \sigma \right) \nonumber\\
    &+ \frac{\epsilon}{2} \left[ k^2 \left( \frac{a(r)}{1-\frac{2M}{r}} + b(r)\left(1-\frac{2M}{r}\right)\right) - b(r)\left( \sigma + \frac{h^2}{r^2} \right)\left(1-\frac{2M}{r}\right)^2\right]\,.
\end{align}

All desired observables can be derived from the equations
\begin{align}
    \dot r 
    &=  \sqrt{-2 V(r)}\,,\label{eq:dotr}\\
    \left(\frac{dr}{d\phi}\right)^2
    &= \frac{\dot r^2}{\dot \phi^2} = - \frac{2 V(r) r^4}{h^2}\label{eq:rphi}\,,\\
    \frac{dt}{dr} 
    &= \frac{\dot t}{\dot r} =  \frac{k}{A\sqrt{-2 V(r)}}\label{eq:tr}\,,\\
    \frac{d\phi}{dr} 
    &= \frac{\dot \phi}{\dot r} =  \frac{h}{\sqrt{-2 V(r)} r^2}\label{eq:phir}\,.
\end{align}
Having introduced all these quantities, we can now compute different observables for the perturbed solution found.

%%%%%%%%%%%%%%%%%%%%%%%%%%%%%%%%%%%%%%%%%%%%%%%%%%%%%%%%%%%%%%%%%%%%%%%%%%%%%%%%%%%%%%%%%%%%%%%%%%%%%
\subsection{Photon sphere}\label{ssec:phsph}
To determine the photon sphere we derive the circular orbits for massless particles ($\sigma=0$). They are characterized by setting $\dot r = 0$ in \eqref{eq:dotr}, hence $V(r)$ and $V'(r)$ must vanish. For the calculation we expand the radius of interest $r_c$ into $r_c= r_0 + \epsilon r_1$, as well as $h=h_0 + \epsilon h_1$ and $k= k_0 + \epsilon k_1$. Solving the equations order by order yields to zeroth order
\begin{align}
	r_0 = 3M,\quad h_{0\pm} = \pm 3 \sqrt{3} k_0 M\,,
\end{align} 
and the first order correction
\begin{align}
    r_1
    &=\frac{\left(229 \sqrt{3}-177\right) \alpha_1+\left(133 \sqrt{3}-1809\right) \alpha_2+2 \left(61 \sqrt{3}+87\right) \alpha_3}{240 \left(\sqrt{3}+3\right) M}+\frac{9 \log (3) (\alpha_1+9 \alpha_2-2 \alpha_3)}{64 M}\\
    &\approx \frac{0.34789 \alpha_1}{M}+\frac{0.000409083 \alpha_2}{M}+\frac{0.0302887 \alpha_3}{M}\,, \label{eq:r1}\\ 
    h_{1\pm}
    &=\pm\frac{3 k_0 \left(\sqrt{3} \alpha_1+\alpha_1-71 \alpha_2+41 \sqrt{3} \alpha_2-10 \alpha_3+6 \sqrt{3} \alpha_3\right)}{8 \left(\sqrt{3}+3\right) M}\pm\frac{9 \left(\sqrt{3}+1\right) k_1 M}{\sqrt{3}+3}\\\
    &\approx \pm \frac{k_0 (0.216506 \alpha_1+0.00111604 \alpha_2+0.0310889 \alpha_3)}{M}\pm 5.19615 k_1 M\,.
\end{align}
It is interesting to see that there exists a two parameter family of teleparallel modifications of general relativity for which $r_1$ vanishes, since one can solve $r_1=0$ for one of the coupling constants $\alpha_I$ in terms of the others.

The obtained result reproduces the findings of~\cite{Bahamonde:2019zea} by setting $\alpha_1=\alpha_2=-\alpha_3=(4/9)\alpha$ which is the $f(\mathbb{T})$-squared power law case.

%%%%%%%%%%%%%%%%%%%%%%%%%%%%%%%%%%%%%%%%%%%%%%%%%%%%%%%%%%%%%%%%%%%%%%%%%%%%%%%%%%%%%%%%%%%%%%%%%%%%%
\subsection{Perihelion Shift}\label{ssec:peri}
Let $r_c$ be a circular orbit of a massive object ($\sigma=1$). The perihelion shift of a nearly circular orbit $r(\phi) = r_c + r_\phi(\phi)$ is determined from \eqref{eq:rphi}, by expanding it in orders of $r_\phi/r_c$ up to second order, and studying the resulting wave equation.

The expression for the perihelion shift is
\begin{align}\label{eq:perihel}
    \Delta \phi =2\pi\Big(\frac{1}{K}-1\Big) =2\pi \left(\frac{h}{r_c^2\sqrt{V''(r_c)}}-1\right)\,.
\end{align}

The effective potential depends on the conserved quantities $k = k_0 + \epsilon k_1$ and $h = h_0 + \epsilon h_1$. For circular orbits $r_c$ the effective potential satisfies, as already mentioned, $V(r_c) = 0$ and $V'(r_c)=0$. These equations can be solved order by order for the energy and angular momentum parameters
\begin{align}
    h_0(r_c) =\pm\frac{r_c}{\sqrt{r_c/M-3}} ,\quad k_0(r_c) = \pm\frac{ r_c-2 M}{\sqrt{r_0(r_c-3 M)}}\,,
\end{align}
as well as for
\begin{eqnarray}
    h_1(r_c) &=&\pm \frac{\Big(r_c\left(3 \mu_c^2-1\right)^{3}(r_c-\mu_0^2 r_c)\Big)^{-1/2}}{320 \mu_c^6 \left(\mu_c^2-1\right)}\Big[-1215 \mu_c^{10} (\alpha_1+\alpha_2+2 \alpha_3)+2160 \mu_c^9 (\alpha_1+2 \alpha_2+3 \alpha_3)\nonumber\\
    &&+90 \mu_c^8 (9 \alpha_1-39 \alpha_2-22 \alpha_3)-2592 \mu_c^7 (\alpha_1+2 \alpha_2+3 \alpha_3)+40 \mu_c^6 (11 \alpha_1+293 \alpha_2+169 \alpha_3)\nonumber\\
    &&+180 \mu_c^4 (\alpha_1+9 \alpha_2-2 \alpha_3)+160 \mu_c^3 (\alpha_1-2 \alpha_2-\alpha_3)+\mu_c^2 \Big(120 \log (\mu_c) (\alpha_1+9 \alpha_2-2 \alpha_3)\nonumber\\
    &&+327 \alpha_1+119 \alpha_2-394 \alpha_3\Big)-240 \mu_c (\alpha_1-2 \alpha_2-\alpha_3)-30 (\alpha_1+\alpha_2+2 \alpha_3)-320 \mu_c^5 (25 \alpha_2+\alpha_3)\Big] \,,
\end{eqnarray}
where $\mu^2_c=1-2M/r_c$. The quantity $k_1$ does not play any role here. Plugging these into \eqref{eq:perihel} yield the perihelion shift as function of $r_c$
\begin{eqnarray}\label{eqn:perihelionshift}
    \Delta\phi(h_{0+},h_{1+}) &=& \Delta\phi_{\rm GR}+ \epsilon\, \Delta\phi_{\epsilon} \\
    &=&6 \pi q + 27 \pi q^2%+135 \pi  q^3 \nonumber
     + \epsilon\,\frac{2 \pi  q^2 }{r_c^2} (16 \alpha_1+\alpha_2+8 \alpha_3)+ \mathcal{O}(q^3)\,,\label{eq:perishift}
\end{eqnarray}
where $q=M/r_c$. It is not surprising that also for this observable there exists a two parameter family of non-trivial teleparallel modification of general relativity, which makes this correction vanish. For example one could choose  $\alpha_2= -8 (2 \alpha_1+\alpha_3)$.

Again, this result is consistent with the $f(\mathbb{T})$-squared power law case reported in~\cite{Bahamonde:2019zea,Ilijic:2018ulf} which is recovered by setting $\alpha_1=\alpha_2=-\alpha_3=(4/9)\alpha$, yielding $\Delta\phi_{\epsilon}=8 \pi  \alpha  q^2/r_c^2$. 

%%%%%%%%%%%%%%%%%%%%%%%%%%%%%%%%%%%%%%%%%%%%%%%%%%%%%%%%%%%%%%%%%%%%%%%%%%%%%%%%%%%%%%%%%%%%%%%%%%%%%
\subsection{Shapiro Delay}\label{ssec:shap}
For the Shapiro delay one considers an emitter of a light pulse at a given radius $r_e$, the light ray propagates through spacetime to a point of closest encounter to the central mass $r_0$ and from there to a mirror at radius $r_m$ where it gets reflected and returns on the same path to the emitter. The Shapiro delay is the difference in the time of flight of the signal in the presence and absence of the gravitational field.

The time required from a radial signal to travel from $r_0$ to another radius $r$ can be obtained by integrating Eq.~\eqref{eq:dotr}, giving us
\begin{equation}\label{time}
t(r,r_0)=\int\limits_{r_0}^r d\bar{r} \sqrt{-2V(\bar{r})}=\int\limits_{r_0}^r d\bar{r}\Big[\left(1-\frac{r_0^2 A(r_0)}{\bar{r}^2 A(\bar{r})}\right)\frac{A(\bar{r})}{B(\bar{r})}\Big]^{-1/2}\,,
\end{equation}
where $\sigma=0$ and
\begin{equation}
   r_0^2=\Big(\frac{h}{k}\Big)^2A(r_0)\,.
\end{equation}
Now, we need to replace the effective potential~\eqref{eq:pot1} and the solution~\eqref{sola}-\eqref{solb} and expand up to first order in $\epsilon$. Doing this, the integrand becomes
\begin{eqnarray}
\Big[\left(1-\frac{r_0^2 \mathcal{A}(r_0)}{\bar{r}^2 \mathcal{A}(\bar{r})}\right)\frac{\mathcal{A}(\bar{r})}{\mathcal{B}(\bar{r})}\Big]^{-1/2}&=&\frac{\mu_0 \left(\mu_0^2-1\right) }{\bar{\mu}^2}\Big(\mu_0^6-2 \mu_0^4+\mu_0^2-\bar{\mu}^2 \left(\bar{\mu}^2-1\right)^2\Big)^{-1/2}- \epsilon  \,\frac{\left(\mu_0^2-1\right)}{2 \mu_0 \bar{\mu}^4}\Big(\mu_0^6-2 \mu_0^4+\mu_0^2\nonumber\\
&&-\bar{\mu}^2 \left(\bar{\mu}^2-1\right)^2\Big)^{-3/2}\times \Big[\bar{\mu}^4 \left(\bar{\mu}^2-1\right)^2 a(r_0)+\mu_0^2 \left(\mu_0^6-2 \mu_0^4+\mu_0^2-2 \bar{\mu}^2 \left(\bar{\mu}^2-1\right)^2\right) a(\bar{r})\nonumber\\
&&-\mu_0^2 \bar{\mu}^4 \left(\mu_0^6-2 \mu_0^4+\mu_0^2-\bar{\mu}^2 \left(\bar{\mu}^2-1\right)^2\right) b(\bar{r})\Big]\,,\label{123}
\end{eqnarray}
where $\bar{\mu}^2=1-2M/\bar{r}$ and $\mu_0^2=1-2M/r_0$. The first term corresponds to the standard GR contribution while the second one will be related to our perturbed solution. The retardation of light (Shapiro delay) will be then defined as
\begin{equation}
  \Delta  t_{\rm Shapiro}(r_{\rm e},r_{\rm m},r_0)=\frac{1}{2}\left(t(r_{\rm e},r_0)+t(r_{\rm m},r_0)-\sqrt{r_{\rm e}^2-r_0^2}-\sqrt{r_{\rm m}^2-r_0^2} \right)\,.
\end{equation}
To be able to integrate Eq.~\eqref{time}, one needs to make some approximations. We assume a small Schwarzschild radius, resp. a small mass of the central object $r \gg M$, and make a power series expansion in the mass parameter, to get the Shapiro delay corresponding to the solution~\eqref{sola}-\eqref{solb}. Approximately the time delay between $r$ and $r_0$ can be written as follows
\begin{eqnarray}
    \Delta t_{\rm Shapiro}(r,r,r_0)&=& \Delta t_{\rm Shapiro,GR}(r,r,r_0)+\epsilon\,  \Delta t_{\rm Shapiro}(r,r,r_0)\\
    &\approx& \frac{M \sqrt{r^2-r_0^2}}{r+r_0}+2 M \log \left(\frac{\sqrt{r^2-r_0^2}+r}{r_0}\right)+\epsilon\,\frac{M^3 \sqrt{r^2-r_0^2}}{10(r+r_0)} \Big[\frac{1}{r_0^2r^2}\Big(24\alpha_1-\alpha_2+2\alpha_3\Big)\nonumber\\
    &&+\frac{2}{r_0^3r}\Big(44 \alpha_1-  \alpha_2+ 7 \alpha_3\Big)+\frac{20}{r_0^4}\Big(4 \alpha_1+ \alpha_3\Big)\Big]\,.\label{eq:shap}
\end{eqnarray}
For the $f(\mathbb{T})=\mathbb{T}+\frac{1}{2}\epsilon\, \alpha \mathbb{T}^2$ case ($\alpha_1=\alpha_2=-\alpha_3=(4/9)\alpha$), we find 
\begin{eqnarray}
    \Delta t_{\rm Shapiro}(r,r,r_0)&=& \Delta t_{\rm Shapiro,GR}(r,r,r_0)+\epsilon\,  \Delta t_{\rm Shapiro}(r,r,r_0)\\
    &\approx& \frac{M \sqrt{r^2-r_0^2}}{r+r_0}+2 M \log \left(\frac{\sqrt{r^2-r_0^2}+r}{r_0}\right)+\epsilon\,\frac{4\alpha M^3 \sqrt{r^2-r_0^2}}{30(r+r_0)} \Big[\frac{7}{ r^3 r_0}+\frac{7}{ r^2 r_0^2}+\frac{17}{ r r_0^3}+\frac{20}{ r_0^4}\Big]\,,\label{eq:shap2}
\end{eqnarray}
which matches the result found in~\cite{Bahamonde:2020bbc} by adding the approximation $r\gg 1$ (which for example, it is valid in the Solar System), which gives us $t_{ \rm Shapiro,\epsilon}(r_,r_0)\approx 8 \alpha  M^3/(3 r_0^4)$. It is also interesting to mention that for the theory $\alpha_2=16 \alpha_1$ and $\alpha_3= -4 \alpha_1$, the contribution coming from the modification drops out and the Shapiro delay is exactly the same as the one predicted by GR. In contrast to the previous observables one has to fix to of the three model parameters to achieve this.

%%%%%%%%%%%%%%%%%%%%%%%%%%%%%%%%%%%%%%%%%%%%%%%%%%%%%%%%%%%%%%%%%%%%%%%%%%%%%%%%%%%%%%%%%%%%%%%%%%%%%
\subsection{Light Deflection}\label{ssec:light}
The light deflection angle $\Delta\varphi$ can be found by integrating~\eqref{eq:phir}, setting $r_0$ as the minimal distance $\dot{r}(r_0)=0$ ($V(r_0)=0$) and replacing
\begin{equation}
   r_0^2=\Big(\frac{h}{k}\Big)^2A(r_0)\,,
\end{equation}
which yields the following expression for $\Delta\varphi$,
%\begin{equation}
 %   \frac{d\phi}{dr}=\frac{\dot{\phi}}{\dot{r}}=\pm\frac{h}{r^2\sqrt{-2V(r)}}=\pm\frac{B^{1/2}}{r^2}\Big(\frac{A(r_0)}{r_0^2A}-\frac{1}{r^2}\Big)^{-1/2}\,.
%\end{equation}
\begin{equation}
    \pi + \Delta\varphi=\pm 2 \int_{r_0}^\infty d\bar r \frac{h}{\bar r^2\sqrt{-2V(\bar r)}}
    = \pm 2 \int_{r_0}^\infty d\bar r \frac{B(\bar r)^{1/2}}{\bar r^2}\Big(\frac{A(r_0)}{r_0^2 A(\bar r)}-\frac{1}{\bar r^2}\Big)^{-1/2}\,.\label{phiis}
\end{equation}
Without loss of generality we choose the $+$ sign when further evaluating \eqref{phiis}. Pictorially this can be understood as the angle which characterizes the difference between the light trajectory at infinity, with or without the presence of a gravitating central mass.

If we replace the metric functions as~\eqref{expansion} in ~\eqref{phiis}, and expand up to first order in $\epsilon$, its integrand becomes
\begin{equation}
    \frac{B(\bar r)^{1/2}}{\bar r^2}\Big(\frac{A(r_0)}{r_0^2 A(\bar r)}-\frac{1}{\bar r^2}\Big)^{-1/2}= \frac{ 4 M \left(r_0^3-\bar{r}^3\right)-2 r_0^3 \bar{r}+2 r_0 \bar{r}^3- \epsilon\,  r_0 \bar{r}^3 a(r_0)}{2 \bar{r}^2 \sqrt{\frac{2 M}{r_0}\left(\frac{r_0^3}{\bar{r}^3}-1\right)-\frac{r_0^2}{\bar{r}^2}+1} \left(2 M \left(r_0^3-\bar{r}^3\right)-r_0^3 \bar{r}+r_0 \bar{r}^3\right)}\,.
\end{equation}
As we pointed out in the previous section, to integrate~\eqref{phiis} we must assume an approximation. Thus, if we replace our solution and perform the approximation $r\gg M$, one finds that the deflection of light will be approximately given by 
\begin{equation}
     \Delta \varphi\approx \varphi_{\rm GR}+\epsilon \, \varphi_{\epsilon} = \frac{4 M}{r_0}+\frac{ M^2}{r_0^2}\left(\frac{15 \pi }{4}-4\right)+\frac{ M^3}{r_0^3}\Big(\frac{244-45 \pi}{6}\Big)+\epsilon\, M^3 \frac{16 (4 \alpha_1+\alpha_3)}{5 r_0^5}%-\frac{\alpha_2}{5 r r_0^4}
     +\mathcal{O}(M^4/r_0^4)\,.
\end{equation}
The first three terms on the most right hand side of the above expression are the well-known GR result expanded up to $M^3/r_0^3$ whereas the fourth one is the modification caused by the teleparallel correction. One can notice that the leading term correction appears with $M^3/r_0^3$. The parameter $\alpha_2$ does not play a role here to leading order, but only to subleading orders. For the $f(\mathbb{T})=\mathbb{T}+\frac{1}{2}\epsilon\, \alpha \mathbb{T}^2$ case, the deflection of light becomes
\begin{equation}
     \Delta \varphi\approx \varphi_{\rm GR}+\epsilon \, \varphi_{\epsilon} = \frac{4 M}{r_0}+\frac{ M^2}{r_0^2}\left(\frac{15 \pi }{4}-4\right)+\frac{ M^3}{r_0^3}\Big(\frac{244-45 \pi}{6}\Big)+\epsilon\, M^3 \frac{64 \alpha}{45 r_0^5}%-\frac{\alpha_2}{5 r r_0^4}
     +\mathcal{O}(M^4/r_0^4)\,,
\end{equation}
which matches the result obtained in~\cite{Bahamonde:2020bbc}.

%%%%%%%%%%%%%%%%%%%%%%%%%%%%%%%%%%%%%%%%%%%%%%%%%%%%%%%
\section{Conclusion}\label{sec:conclusion}
Spherically symmetric vacuum solutions form a very important foundation for the viability of theories of gravity. In first approximation they describe planetary systems as the solar system, they describe non-rotating  black holes and are the first step towards realistic rotating black hole solutions, they can be used as exterior solutions for stars, as well as for the derivation of quasi-normal modes after a merger event. 

In this article we derived the most general spherically symmetric field equations for a teleparallel theory of gravity. Moreover, we solved the field equations for two teleparallel theories of gravity explicitly. First, we had a quick look at conformal teleparallel gravity in spherical symmetry, which turned out not be a predictive theory. Second, we solved the field equations perturbatively for the most general quadratic polynomial teleparallel perturbation of GR. The theory contains three free parameters $\alpha_i, i=1,2,3$ which can be constrained by experiments. We explicitly derived how these parameters influence the photon sphere, the perihelion shift, the Shapiro delay and the light deflection. 

For the photon sphere we found that its acquires a teleparallel correction of the form $M^{-1}\sum_{i=1}^3 a_i \alpha_i$, where the values of the $a_i$ are fixed, see \eqref{eq:r1}. Hence, the parameters can be chosen such that a larger, smaller or identical shadow of a black hole compared to general relativity is predicted. Fixing the photon sphere to an observed value would fix one of the three parameters in terms of the other two.

A similar observation can be made for the perihelion shift \eqref{eq:perishift}. The teleparallel influence is determined by a different linear combination of the model parameters as $\frac{2\pi q^2}{r_c^2}\sum_{i=1}^3 b_i \alpha_i$, where the values of the $b_i$ are fixed. In principle, larger, smaller or an equal value to the prediction of general relativity is possible. Fixing the perihelion shift prediction to an observed value determines one of the three parameters in terms of the other, but with a different relation as for the photon sphere. Hence these two observables together constrain two of three model parameters.

For the Shapiro delay the situation is different. We found that it depends on the teleparallel model parameters in a more involved way. Schematically the correction to the GR prediction is given by $ M^3\frac{\sqrt{r^2-r_0^2}}{10 (r+r_0)} [(r_0 r)^{-2} \sum_{i=1}^3 c_i \alpha_i + r_0^{-3} r^{-1} \sum_{i=1}^3 d_i \alpha_i + r_0^{-4} (4 \alpha_1 + \alpha_3) ]$, see \eqref{eq:shap}. Due to the appearance of the different powers of the radii $r$ and $r_0$, between which one measures the time delay, measurements for different values of $r$ constrain  the three different combinations of the parameters and thus the Shapiro delay is way more sensitive to detect or constrain teleparallel corrections to GR as the previously discussed photon sphere and perihelion shift. Combining the Shapiro delay with one of the other observables suffices to constrain all three parameters of the theory.

Last but not least we studied the teleparallel correction to the light deflection. Most interestingly, to leading order it only depends on the parameter combination $4\alpha_1 + \alpha_3$, which also appears in the Shapiro delay. 
We conclude that a combined high precision measurement of the Shapiro delay and light deflection presents a very good ground to test teleparallel modifications of GR against experimental observations. Considering in addition perihelion shift observations it is possible to constrain all three parameters of theory. The photons sphere is not a direct observable and the black hole shadow detection has not yet reached the precision to yield additional constraints to the aforementioned ones.

Future research, which will rely on the spherically symmetric solutions presented here, are the calculation of the quasi normal modes spectrum of black holes and neutrons stars, that are formed in a collision of a binary system. This will predict the influence of teleparallel modifications of GR in gravitational wave signals. From the more theoretical aspects, the analysis of quasi normal modes, starting from spherically symmetric solutions, will additionally provide new insights in the long standing question of the number of degrees of freedom of $f(\mathbb{T})$ gravity. Moreover, the influence of teleparallel gravity on stars can be studied by searching for interior solutions which match the presented vacuum solutions.

%%%%%%%%%%%%%%%%%%%%%%%%%%%%%%%%%%%%%%%%%%%%%%%%%%%%%%%
\begin{acknowledgments}
CP was supported by the Estonian Ministry for Education and Science through the Personal Research Funding Grants PSG489 and the European Regional Development Fund through the Center of Excellence TK133 ``The Dark Side of the Universe''. SB is supported by Mobilitas Pluss N$^\circ$ MOBJD423 by the Estonian government. 
\end{acknowledgments}

\appendix

\section{Scalars in spherical symmetric}\label{appendixB}
The scalars constructed by contraction of the decomposition of torsion in spherical symmetry for the most general tetrad~\eqref{sphtetrad} with the condition~\eqref{eq:coordcond1} and the equation~\eqref{eq:coordcond2} are
\begin{eqnarray}
    T_{\rm ten}&=&\Big[C^3 C_4^2 \left(C^2-C_5^2\right) \left(C_1^2-C_3^2\right)^2\Big]^{-1}\times\Big[-C^3 \Big(-C_1^2 \left(C_3^2 \left(C'^2+2 C_4 C_5'+2 C_4^2+C_5'^2\right)+C_5^2 \left(C_1'^2-C_3'^2\right)\right)\nonumber\\
    &&+C_1^4 \left(C'^2+2 C_4 C_5'+C_4^2+C_5'^2\right)-2 C_1 C_3^2 C_4 C_5 C_1'+2 C_1^3 C_4 C_5 C_1'+C_3^4 C_4^2\Big)\nonumber\\
    &&+2 C^2 C_1^2 C_5 C' \left(-C_1 C_5 C_1'+C_1^2 \left(2 C_4+C_5'\right)+C_3 \left(C_5 C_3'-C_3 \left(2 C_4+C_5'\right)\right)\right)+2 C^4 C_1^2 C' \left(C_1 C_1'-C_3 C_3'\right)\nonumber\\
    &&+4 C_1^2 C_4 C_5^3 C' \left(C_3^2-C_1^2\right)+C^5 C_1^2 \left(C_3'^2-C_1'^2\right)+C C_4 C_5^2 \left(C_1^2-C_3^2\right) \left(2 C_1 C_5 C_1'+C_1^2 \left(C_4+2 C_5'\right)-C_3^2 C_4\right)\Big]\,,\label{Tscalar1}\\ \nonumber \\
    T_{\rm vec}&=&\frac{1}{C_1^2-C_3^2}\Big[\left(\frac{2 C_3 C_5}{C^2}+\frac{C_1 \left(C_1 C_3'-C_3 C_1'\right)}{C_4 \left(C_1^2-C_3^2\right)}\right)^2-\frac{C_1^2 \left(\frac{2 C_4 C_5-2 C C'}{C^2}+\frac{C_3 C_3'-C_1 C_1'}{C_1^2-C_3^2}\right)^2}{C_4^2}\Big]\,,\label{Tscalar2}\\  \nonumber \\
    T_{\rm ax}&=&\Big[9 C^4 C_4^2 \left(C^2-C_5^2\right) \left(C_1^2-C_3^2\right)\Big]^{-1}\times\Big[4 \Big(C^2 C_5^2 \left(C_1^2 \left(C'^2+4 C_4 C_5'-8 C_4^2\right)+8 C_3^2 C_4^2\right)+2 C^3 C_1^2 C_5 C' \left(2 C_4-C_5'\right)\nonumber\\
    &&-4 C C_1^2 C_4 C_5^3 C'+C^4 \left(C_1^2 \left(C_5'-2 C_4\right)^2-4 C_3^2 C_4^2\right)+4 C_4^2 C_5^4 \left(C_1^2-C_3^2\right)\Big)\Big]\,,\label{Tscalar3}\\  \nonumber \\
    P_1&=&2 \sin\theta\Big[3 C C_4 \sqrt{C^2-C_5^2} \left(C_1^2-C_3^2\right)\Big]^{-1}\times\Big[C^2 C' \left(-C_1 C_3 C_5 C_1'+C_1^2 \left(C_5 C_3'+4 C_3 C_4\right)-4 C_3^3 C_4\right)\nonumber\\
    &&+2 C_3 C_4 C_5^2 C' \left(C_3^2-C_1^2\right)+C^3 \left(C_1 C_3 C_1' C_5'+C_1^2 C_3' \left(2 C_4-C_5'\right)-2 C_3^2 C_4 C_3'\right)\nonumber\\
    &&-2 C C_4 C_5 \left(C_1^2-C_3^2\right) \left(C_5 C_3'+C_3 C_5'\right))\Big]\,,\label{Tscalar4}\\  \nonumber \\
    P_2&=& 3C_1^2\csc\theta \Big[C^5 C_4^3 \sqrt{C^2-C_5^2} \left(C_1^2-C_3^2\right)^3\Big]^{-1}\times \Big[C^2 C' \left(C_1 C_3 C_5 C_1'-C_1^2 \left(C_5 C_3'+C_3 C_4\right)+C_3^3 C_4\right)\nonumber\\
    &&+2 C_3 C_4 C_5^2 C' \left(C_1^2-C_3^2\right)+C^3 \left(-C_1 C_3 C_1' C_5'+C_1^2 C_3' \left(C_4+C_5'\right)-C_3^2 C_4 C_3'\right)\nonumber\\
    &&-C C_4 C_5 \left(C_1^2-C_3^2\right) \left(C_5 C_3'+C_3 C_5'\right)\Big]\,.\label{Tscalar5}
\end{eqnarray}
\section{Antisymmetric field equations}\label{appendixC}
The functions $Q_i$ appearing in the antisymmetric field equations~\eqref{anti1}-\eqref{anti2} are
\begin{eqnarray}
    Q_1&=&\Big[C^3 C_4^2 \left(C_1^2-C_3^2\right)^3\Big]^{-1}\times\Big[-C_1^2 C_3 \Big(C^2 C_3 \left(C C_3' C_4'-C_4 \left(2 C' C_3'+C C_3''\right)\right)+4 C_3^2 C_4^2 \left(2 C_5 C'-C C_5'\right)\nonumber\\
    &&+C^3 C_4 \left(C_1'^2+2 C_3'^2\right)\Big)+C_3^3 C_4 \left(2 C_3^2 C_4 \left(2 C_5 C'-C C_5'\right)-C^3 C_1'^2\right)+C^2 C_1^3 \Big(C_3 \left(2 C_4 C' C_1'+C C_4 C_1''-C C_1' C_4'\right)\nonumber\\
    &&+C C_4 C_1' C_3'\Big)+C^2 C_1 C_3^2 \left(C_3 \left(C C_1' C_4'-C_4 \left(2 C' C_1'+C C_1''\right)\right)+3 C C_4 C_1' C_3'\right)\nonumber\\
    &&+C_1^4 \left(C^2 \left(C C_3' C_4'-C_4 \left(2 C' C_3'+C C_3''\right)\right)+2 C_3 C_4^2 \left(2 C_5 C'-C C_5'\right)\right)\Big]\,,\\
      Q_2&=&\frac{8 C_3 \left(C_5 C'-C C_5'\right)}{9 C^3 \left(C_1^2-C_3^2\right)}\,,\\
        Q_3&=&\frac{C^2 C_1 C_3 C_1'-C_1^2 \left(C^2 C_3'+2 C_3 C_4 C_5\right)+2 C_3^3 C_4 C_5}{C^2 C_4 \left(C_1^2-C_3^2\right)^2}\,,\\
          Q_4&=&-\frac{\sqrt{C^2-C_5^2} \left(C_1^2 \left(2 C C'-2 C_4 C_5\right)+C^2 C_1 C_1'+2 C_3^2 C_4 C_5\right)}{C^4 C_4 \left(C_1^2-C_3^2\right)}\,,\\
            Q_5&=&2\Big[9 C^4 C_4^3 \left(C^2-C_5^2\right)^{3/2} \left(C_1^2-C_3^2\right)\Big]^{-1}\times\Big[C^2 C_4 C_5^3 \left(C_1^2 \left(C'^2-8 C_4^2\right)-2 C_1 C_4 C_5 C_1'+8 C_3^2 C_4^2\right)-2 C C_1^2 C_4^2 C_5^4 C'\nonumber\\
            &&+C^3 C_1 C_5^2 \left(C_4 C_5 C' C_1'+C_1 \left(C_4 C_5 C''-C_5 C' C_4'-2 C_4 C' C_5'+4 C_4^2 C'\right)\right)-C^5 C_1 \Big(C_4 C_5 C' C_1'\nonumber\\
            &&+C_1 \left(C_4 C_5 C''-C_5 C' C_4'+2 C_4^2 C'\right)\Big)+C^4 C_5 \Big(C_1 C_4 C_5 C_1' \left(4 C_4-C_5'\right)+C_1^2 \left(C_5 C_4' C_5'+C_4 \left(C_5'^2-C_5 C_5''\right)+4 C_4^3\right)\nonumber\\
            &&-4 C_3^2 C_4^3\Big)+C^6 C_1 \left(C_4 \left(C_1' C_5'+C_1 C_5''\right)-2 C_4^2 C_1'-C_1 C_4' C_5'\right)+4 C_4^3 C_5^5 \left(C_1^2-C_3^2\right)\Big]\,,\\
              Q_6&=&\frac{2 C_1^2 \left(-C C_5 C'+C^2 \left(C_5'-2 C_4\right)+2 C_4 C_5^2\right)}{9 C^2 C_4^2 \sqrt{C^2-C_5^2} \left(C_1^2-C_3^2\right)}\,.
\end{eqnarray}
%%%%%%%%%%%%%%%%%%%%%%%%%%%%%%%%%%%%%%%%%%%%%%%%%%%%%%%
\bibliographystyle{utphys}
\bibliography{SBHTP}

\end{document}